\documentclass[3p,times,procedia]{elsarticle}
\usepackage{nupha_ecrc}

\volume{00}

\firstpage{1}

\journalname{Nuclear Physics A}

\runauth{}

\jid{nupha}

\jnltitlelogo{Nuclear Physics A}

\usepackage{amssymb}

\usepackage[figuresright]{rotating}

\begin{document}

\begin{frontmatter}



\dochead{}

\title{Charm hadrons above $T_c$\footnote{This work was supported by U.S. Department of Energy under
Contract No. DE-SC0012704.}}


\author{Swagato Mukherjee, P\'eter Petreczky and Sayantan Sharma}

\address{Physics Department, Brookhaven National Laboratory, Upton, NY 11973, USA}

\begin{abstract}
From the analysis of the lattice data on fluctuations and correlations
of charm we conclude that open charm
meson and baryon-like excitations exist above the QCD crossover temperature,
and in fact are the dominant degrees of freedom for thermodynamics in the
vicinity of the transition.
Charm quarks become the dominant charm degrees of freedom only for 
temperatures $T>200$ MeV. 
\end{abstract}


\end{frontmatter}





\section{Introduction}
Now it is established that there is a chiral transition temperature in QCD with
the physical quark masses at temperature $T_c=(154 \pm 9)$ MeV \cite{Bazavov:2011nk,Borsanyi:2010bp}.
The chiral crossover temperature is related to the chiral phase transition temperature in the 
limit of vanishing up and down quark masses through the universal $O(4)$ 
scaling \cite{Bazavov:2011nk}. On the other hand 
it is not possible
to define a deconfinement transition temperature in QCD with the physical quark masses.
The Polyakov loop, which is an order parameter for deconfinement is not related
to the singular behavior of the free energy in the massless limit. Alternatively, 
one can use fluctuations and correlations of conserved charges 
\begin{equation}
\chi_{n}^{X}=T^n \frac{\partial^{n} p(T,\mu_X,\mu_Y)}{\partial \mu_X^n},~
\chi_{n m}^{XY}=T^{m+n} \frac{\partial^{m+n} p(T,\mu_X,\mu_Y)}{\partial \mu_X^n \partial \mu_Y^m}
\end{equation}
to study the deconfinement
aspects of the QCD transition \cite{Bazavov:2012jq,Borsanyi:2011sw}.
Here $p$ is the pressure, $\mu_X$ and $\mu_Y$ are the chemical potentials corresponding
to some conserved charges $X$ and $Y$. The quantities defined by Eq. (1) are also
called the diagonal and off-diagonal susceptibilities.
At low temperature fluctuations and correlations
of conserved charges are well described by hadron resonance gas (HRG) model, i.e.
by assuming that thermodynamics can be described by adding contributions from
all the hadrons and hadronic resonances \cite{Bazavov:2012jq,Borsanyi:2011sw}.
It has been shown, however, that for
the strangeness correlation and fluctuations the HRG description breaks down abruptly
above $T_c$ \cite{Bazavov:2013dta}. The situation with fluctuations and correlations
involving charm is similar \cite{Bazavov:2014xya}. In that sense the chiral transition
temperature can be considered as a deconfinement temperature. At sufficiently
high temperature fluctuations and correlations of conserved charges can be 
understood in terms of quark degrees of freedom, i.e. they can be understood in terms
of resummed perturbative approach \cite{Bazavov:2013uja,Ding:2015fca}.
However, just above $T_c$ the nature of the relevant degrees of freedom remains unclear. While vacuum
hadrons are clearly not the right degrees of freedom above $T_c$ it
is unlikely that the thermodynamics can be understood in terms of quark degrees
of freedom only near $T_c$. 

To explain the strongly coupled nature of quark gluon plasma
created in heavy ion collisions it was proposed that bound states of partons can exist
for for not too high temperatures \cite{Shuryak:2003ty}.
Lattice studies, though, remained inconclusive whether such bound states could actually exist 
(see e.g. Ref. \cite{Wetzorke:2001dk}).
An important signature of formation of strongly coupled quark gluon plasma is
the elliptic flow and nuclear modification factor of heavy flavor hadrons. 
Most of the models that try to describe these quantities rely on Langevin dynamics
and energy loss of heavy quarks (see e.g.\cite{Moore:2004tg}).
However, the importance of possible heavy-light (strange) 
bound states near the transition temperature was pointed out in Refs. 
\cite{He:2012df}, in particular for the simultaneous description of 
elliptic flow and nuclear modification factor of $D_s$ mesons \cite{He:2012df}.

In this contribution we try to infer the relevant charm degrees of freedom near
the transition temperature using the lattice data on fluctuation and correlations
involving charm quarks \cite{Bazavov:2014xya}. Further details of this study can be found in
Ref. \cite{Mukherjee:2015mxc}.

\section{Results}
We will use the lattice data on $\chi_{n m}^{Xc}$ and $\chi_n^c$ up to fourth order from Ref. \cite{Bazavov:2014xya}.
Here $X$ is either the baryon number, $X=B$ or light quark number, $X=u$. 
We start our discussions with the off-diagonal susceptibilities which directly probe the 
interaction of charm quarks with the light quarks in the heatbath.
In Fig. 1 we show the light and charm
quark number correlations up to forth order normalized by the charm fluctuations $\chi_2^c$. 
The normalization by $\chi_2^c$ cancels out the trivial quark mass effects. At high temperatures
$\chi_{11}^{uc},~\chi_{13}^{uc}$ and $\chi_{31}^{uc}$ are expected to be small, as these
correlations arise from quark line disconnected diagrams. 
The contribution of these diagrams is strictly speaking non-perturbative
but can be calculated using lattice calculations in the dimensionally reduced effective theory for
high temperature QCD \cite{Hietanen:2008xb}. The remaining off-diagonal susceptibility $\chi_{22}^{uc}$ is much
larger in the high temperature region. This is expected since it arises from the plasmon contribution
to the pressure, $\sim T m_D^3$ and the dependence of the Deybe mass $m_D$ on the quark chemical potentials. 
Since we expect that most of the quark mass effects cancel out by taking the ratio of off-diagonal susceptibilities
to $\chi_2^c$ it is sensible to compare the lattice results on these ratios to the HTL resummed perturbative results
for $\chi_{22}^{ud}/\chi_2^u$ \cite{Haque:2014rua} and the result from dimensionally reduced effective theory for 
$\chi_{11}^{ud}/\chi_2^u$ \cite{Hietanen:2008xb}. This comparison is shown in Fig. 1. As one can see from
the figure the lattice results are in line with weak coupling expectations for $T>200$ MeV. This implies
that for $T>200$ MeV charm quarks are the relevant degrees of freedom. At lower temperatures all the off-diagonal
susceptibilities become large and the weak coupling pattern does not hold. In particular $\chi_{13}^{uc}$ is
only about factor three smaller than  $\chi_{22}^{uc}$. Thus, we may expect that the nature of charm degrees of
freedom changes for $T<200 $ MeV. 
\begin{figure}
\includegraphics[width=7cm]{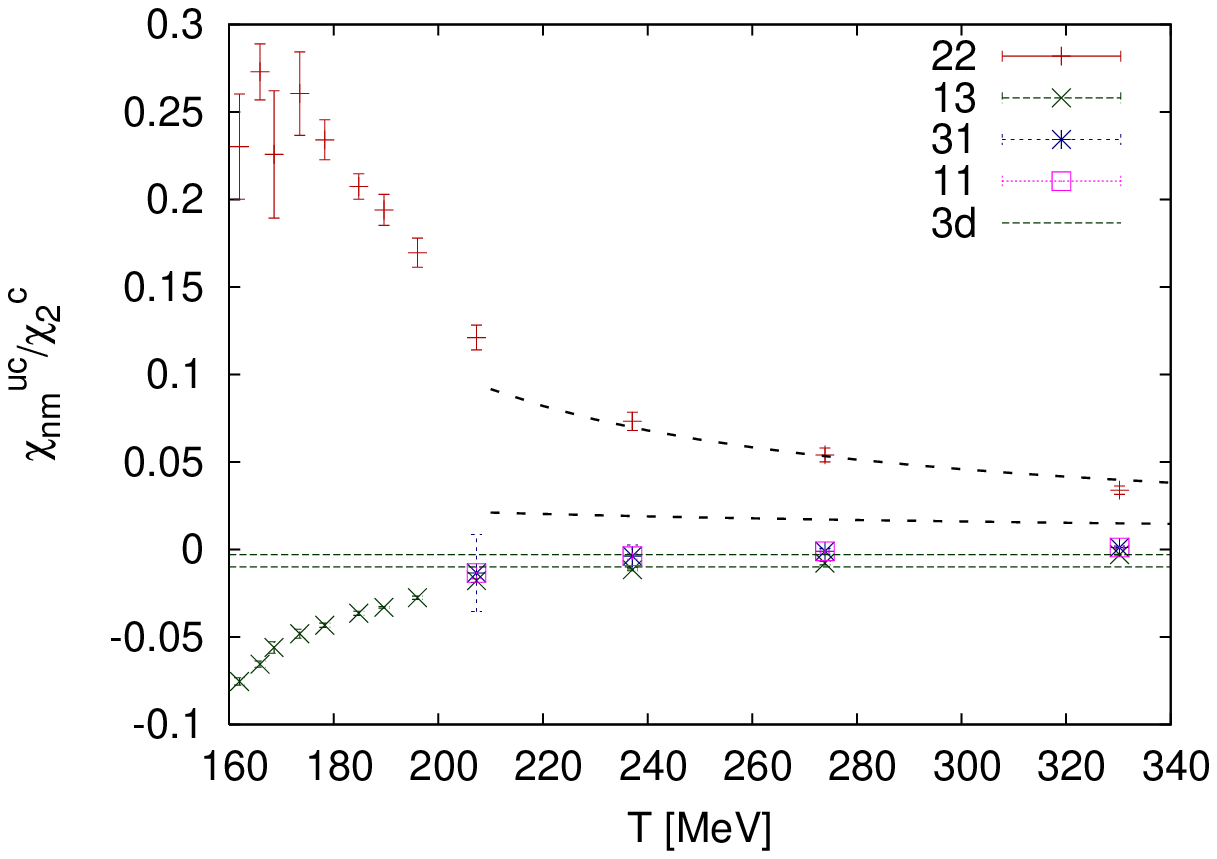}
\includegraphics[width=6cm]{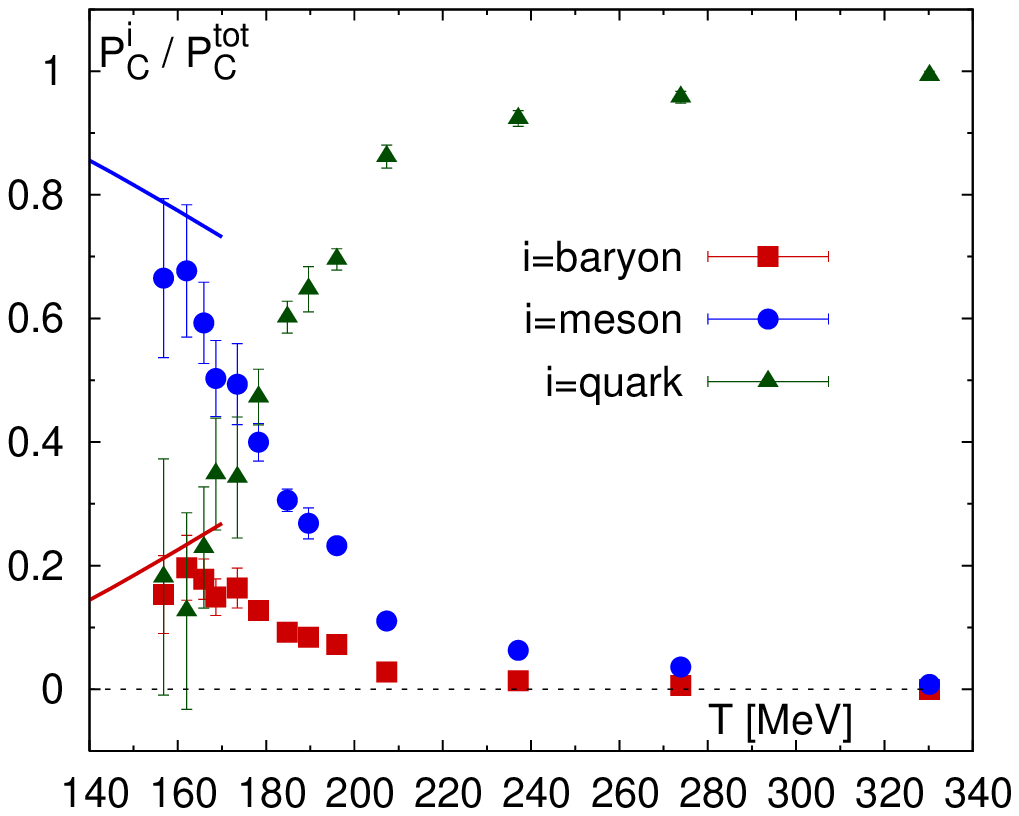}
\caption{The off-diagonal quark number susceptibilities $\chi_{nm}^{uc}$ normalized by charm fluctuations
$\chi_2^c$ (left) and the partial pressures of charm quarks, charm mesons and charm baryons normalized
by the total charm pressure (right).
In the left panel we show the results of the 
resummed
purbation theory (dashed lines) and in dimensionally reduced effective theory (thin dotted lines).
The lines in the left panel correspond to HRG.
}
\end{figure}

Because of the large charm quark mass, excitations that carry charm quantum number are expected
to be a good quasi-particle close to $T_c$, i.e. their widths are small compared to the mass. 
Charm degrees of freedom could be carried by charm quasiquarks with effective mass
that depends on the temperature and the chemical potentials.
However, this alone cannot account for the data on off-diagonal susceptibilities \cite{Mukherjee:2015mxc}.
Therefore, we postulate the existence of charm mesons and baryons above
$T_c$ and we write the partial charm pressure as function of $T$, $\mu_B$ and $\mu_c$
as contributions from charm mesons, charm baryons and charm quarks:
\begin{equation}
p_c(T,\mu_B,\mu_c)=p_M(T) \cosh\left(\frac{\mu_c}{T}\right)+
p_B(T) \cosh\left(\frac{\mu_c+\mu_B}{T}\right)+
p_Q(T) \cosh\left(\frac{\mu_c+\mu_B/3}{T}\right).
\label{pc}
\end{equation}
In the above expression we used the fact that due to the large charm quark mass the Boltzmann approximation is
valid for all charm particles. Furthermore, we neglected the contribution from $|c|=2,3$ sectors as the corresponding
contributions have been shown to be very small \cite{Bazavov:2014xya}. Using the lattice data on correlations and 
fluctuations of charm and light degrees of freedom we have four independent 
quantities, $\chi_2^c$, $\chi_{13}^{Bc}$, $\chi_{31}^{Bc}$ 
and $\chi_{22}^{Bc}$, since $\chi_4^c=\chi_2^c$ and $\chi_{11}^{Bc}=\chi_{13}^{Bc}$ if the $|c|=2,3$ sectors
are neglected.
Taking the appropriate derivatives of $p_c$ with respect to the chemical potentials
one can calculate the correlations and fluctuations involving charm quantum number at $\mu_{c,B}=0$ in terms of the partial 
pressures $p_M(T), p_B(T)$ and $p_Q(T)$ and vice-versa 
re-experess the partial pressures in terms of the above susceptibilities.
We use the lattice data on $\chi_2^c$, $\chi_{13}^{Bc}$ and $\chi_{22}^{Bc}$ to obtain the values
of $p_M(T), p_B(T)$ and $p_Q(T)$.
The three different partial pressures in our model normalized by the total charm pressure 
are shown in Fig. 1. The charm meson and baryon pressures are significantly larger than zero till $T\sim 200$ MeV.
In fact, for temperatures close to 
$T_c$ meson and baryon-like excitations provide the dominant contribution to the charm pressure.
At or around $T_c$ the partial meson and baryon pressures agree within errors with the corresponding partial
pressures from Quark Model HRG, i.e. with HRG model that includes additional states from quark model calculations
that are not yet experimentally observed  \cite{Bazavov:2014xya}. For higher temperatures, the partial meson
and baryon pressures slowly decrease. This means that charm baryon and meson like excitations above $T_c$ are
clearly different from the vacuum charm mesons and baryons. This is in agreement with the lattice calculations of
spatial meson correlation functions and spectral functions, which indicate that there are significant in-medium
modifications of charm mesons already at or below $T_c$ \cite{Bazavov:2014cta}.
The partial charm quark pressure is quite small at $T_c$ but rapidly rises with increasing temperature and 
for $T>200$ MeV it is the dominant contribution to the total charm pressure. At the same time the meson
and baryon pressures become very small at $T>200$ MeV. The lattice results for $\chi_{31}^{Bc}$ have not
been used in the above considerations. These data provide an independent constraint on the partial pressures which
can be used to check the consistency of the proposed model. It turns out that within large errors this constraint
is satisfied \cite{Mukherjee:2015mxc}.

One can ask the question why different contribution to the pressure become very small in certain temperature
limits. The key for understanding this is to realize that quasi-particles that build up the partial pressures
have finite width. At the same time the spectral functions corresponding to these quasi-particles should vanish
at energies much smaller than the charm quark mass. 
In other words, charm excitations will appear as broad assymetric peaks in various spectral functions.
A detailed treatment of thermodynamics of quasi-particle with
finite width was developed in Refs. \cite{Biro:2014sfa}. It was shown that broad
asymmetric spectral functions lead to partial pressures that are considerably smaller than those obtained
with zero width quasi-particles of the same mass, and for sufficiently large width the partial pressures can be
made arbitrarily small. Thus, the smallness of the charm quark pressures at $T_c$ implies that charm quarks 
have a large width there, while the width of meson and baryon like excitations increases with the temperatures
and these excitations become very broad for $T>200$ MeV. Thus, the temperature dependence
of the partial pressures shown in Fig. 1 fits the common expectations: charm quarks are not ``good'' quasi-particles
close to $T_c$ due to strongly coupled nature of the deconfined matter and the corresponding spectral functions
have only broad structures. Meson and baryon like excitations are expected to  gradually melt with temperatures, i.e.
the corresponding width increase with the temperature to the point that no structures around the charm quark mass
can be identified in the spectral functions. Such gradual melting has been studied for quarkonium spectral 
functions using potential models (see e.g. Ref. \cite{Petreczky:2010tk}). 
Fig. 1 suggests that melting of meson and baryon like excitations happens gradually, and
complete melting only occurs for $T>200$ MeV. For these temperatures
the absence of any peak like structure in the spectral functions of meson like excitations can be also seen in 
spatial correlation functions calculated on the lattice. The long distance behavior of the spatial meson correlation 
functions is controlled by the meson screening mass. If the in-medium mesons are completely 
melted the corresponding screening mass approaches the free theory value 
$\sqrt{m_{q1}^2+(\pi T)^2}+\sqrt{m_{q2}^2+(\pi T)^2}$, with $m_{q1}$ and $m_{q2}$ being the relevant quark masses.
In the case of 
charmonium this happens at temperatures $T>300$ MeV, while for a correlated pair consisting of a light and a charm 
meson, this happens at $T>250$ MeV \cite{Bazavov:2014cta} which is not inconsistent with the above considerations.

\section{Conclusions}
In summary, we examined the temperature dependence of the diagonal and off-diagonal susceptibilities 
involving charm and light quarks in the temperature range from $T_c$ to about $330$ MeV.
We have found that the lattice results on quark number susceptibilities are compatible with weak
coupling expectations for $T>200$ MeV, while below this temperature the off-diagnoal  susceptibilities
are much larger indicating strong interaction/correlation between charm quarks and light quarks. 
We proposed that the total pressure due to charm degrees of freedom can be considered as sum of 
partial pressures of charm quarks, charm meson and baryon-like excitations. We  estimated the corresponding partial 
pressures from first principles lattice data on fluctuations and correlations of the charm. We found the charm 
meson and baryon pressures are significant above $T_c$ and become negligible only at $T>200$ MeV, 
implying the charm meson and baryon like excitations may exist up this temperature. In fact, in the vicinity of 
the transition temperature meson and baryon like excitations are the dominant ones in terms of thermodynamics, while
the contribution of charm quarks is marginal. This may have important consequences for modeling the
heavy flavor elliptic flow and heavy flavor nuclear modification factor at small and moderate
$p_T$.



\bibliographystyle{elsarticle-num}
\bibliography{ref}







\end{document}